\documentclass[onecollarge,natbib]{svjour2}
\bibpunct{[}{]}{;}{n}{}{,} 
\smartqed  
\usepackage{graphicx}
\usepackage{bm}
\journalname{Few-Body Systems (EFB22)}
\begin{document}

\title{Electroweak structure of light nuclei within chiral effective 
field theory
}


\author{Laura E. Marcucci}


\institute{Laura E. Marcucci \at
           University of Pisa and INFN-Pisa, I-56127 Pisa (Italy) and
           Gran Sasso Science Institute (INFN), I-67100 L'Aquila (Italy) \\
           Tel.: +39-050-2214901\\
           Fax: +39-050-2214887\\
           \email{laura.marcucci@df.unipi.it}           
}

\date{Received: date / Accepted: date}

\maketitle

\begin{abstract}
We review the results of the most recent calculations for
the electromagnetic structure of light nuclei,  
the weak muon capture on deuteron and $^3$He and the weak 
proton-proton capture reaction at energies of astrophysical interest,
performed within the chiral effective field theory framework.
\end{abstract}

\section{Introduction}
\label{intro}

Among the great advantages of the chiral effective field theory 
($\chi$EFT) framework, the two
following ones are here of interest: 
(i) the possibility of deriving nuclear 
electroweak (EW) currents consistently with
the nuclear interaction, and (ii) 
the possibility of setting a hierarchy among the
different contributions, both for the interactions and the currents.
In fact, it is well known
that $\chi$EFT can justify {\it a priori} the empirical observation
that the contribution of 
three-nucleon interactions to nuclear structure is far less significant
than that of the two-nucleon force.
Furthermore, 
the $\chi$EFT power counting
allows to recognize which are the most significant contributions
also among the different currents. 

The idea of using $\chi$EFT to derive the nuclear EW transition
operators was first implemented by Park {\it et al.}~\cite{Par96} in the 
nineties. They derived the nuclear electromagnetic (EM) current and charge 
operators, within the so-called 
heavy-baryon chiral perturbation theory (HB$\chi$PT) approach,
where the baryons are treated as heavy static sources, and
the perturbative 
expansion is performed in terms of the involved momenta over the baryon 
mass. The weak axial 
current and charge operators were derived by the same authors
few years later~\cite{Par03}, 
and applied to weak reactions of astrophysical interest within a 
``hybrid'' approach, in which nuclear wave functions were obtained 
from phenomenological potentials---the $\chi$EFT potentials
available at the time were not yet as accurate as the phenomenological
ones. Only very recently, these $\chi$EFT weak operators have been used 
to study weak processes which involve few-nucleon systems
in conjunction with nowadays accurate $\chi$EFT potentials,
in particular the two-nucleon (NN) potential derived at
next-to-next-to-next-to leading order (N3LO) by Entem and 
Machleidt~\cite{Ent03}, augmented, when needed, by the three-nucleon
interaction (TNI) obtained at next-to-next-to leading order (N2LO), in the 
version of Ref.~\cite{Nav07}. In particular, 
the muon captures on deuteron and $^3$He, in the non-breakup 
channel~\cite{Mar12},
and the proton-proton weak capture 
(the so-called $pp$ reaction), 
in a wide energy range~\cite{Mar13} have been considered. 

Few years ago, the problem of deriving the EM current
and charge operators in $\chi$EFT has been revisited by
Pastore {\it et al.}~\cite{Pastore} and, in parallel, by
K\"olling {\it et al.}~\cite{Koelling}. Pastore {\it et al.} have used 
time-ordered perturbation theory (TOPT) to calculate the
EM transition amplitudes,
which allows for an easier treatment of the so-called
reducible diagrams than the HB$\chi$PT approach. On the other hand,
K\"olling {\it et al.} have used the method of unitary transformation,
the same one used to derive the chiral potentials mentioned above.
We will focus here only on the work of Pastore {\it et al.}, but we would
like to remark that the results obtained by these two groups, although
with different methods, are in good agreement with each other.
The new set of NN EM currents derived by Pastore {\it et al.}
have been found significantly different 
from those of Park {\it et al.}, as it will be discussed 
in the next section, where the $\chi$EFT EW
operators will be reviewed. This contribution then continues with
Sec.~\ref{sec:res}, where we present the results of the most recent 
$\chi$EFT calculations of EW observables, in particular
the EM structure of $A=2,3$ nuclei~\cite{Pia13}, the rate for 
muon captures on deuteron and $^3$He~\cite{Mar12}, and the $pp$
reaction astrophysical $S$-factor~\cite{Mar13}.
Some concluding remarks will
be presented in Sec.~\ref{sec:concl}.

\section{Electroweak transition operators}
\label{sec:ew_op}

The EW transition operators consist of six terms: the EM 
and the weak axial and vector charge and
current operators. The weak vector current and charge operators can be 
related to the corresponding EM ones applying the conserved-vector-current (CVC)
hypothesis, which basically links these via
a rotation in the isospin space. Therefore, we choose to consider the
EM and axial operators. In this contribution, we will limit
ourselves to the current operators, due to limitation of space.
Both $\chi$EFT 
current operators can be expanded in powers of pions' and nucleons' 
momenta, $Q$, and consist of long- and intermediate-range components which
are described in terms of one- and two-pion exchange contributions, 
as well as contact currents
which encode the short-range physics. These last operators involve a number 
of so-called low-energy constants (LECs), to be fixed to 
experimental data. 

\begin{figure}
  \includegraphics[width=8cm,height=4cm]{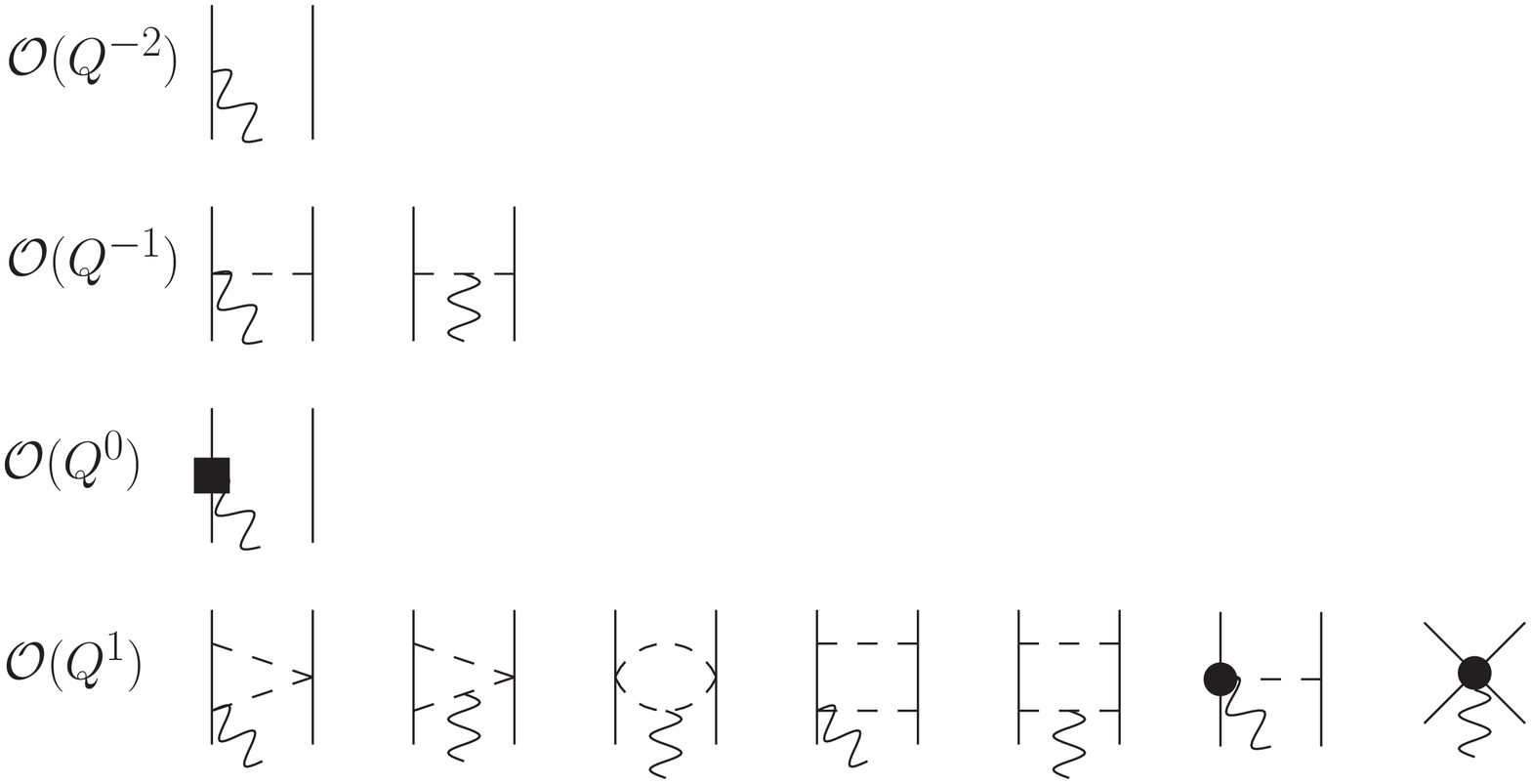}
\hspace*{2cm}
  \includegraphics[width=0.2\textwidth]{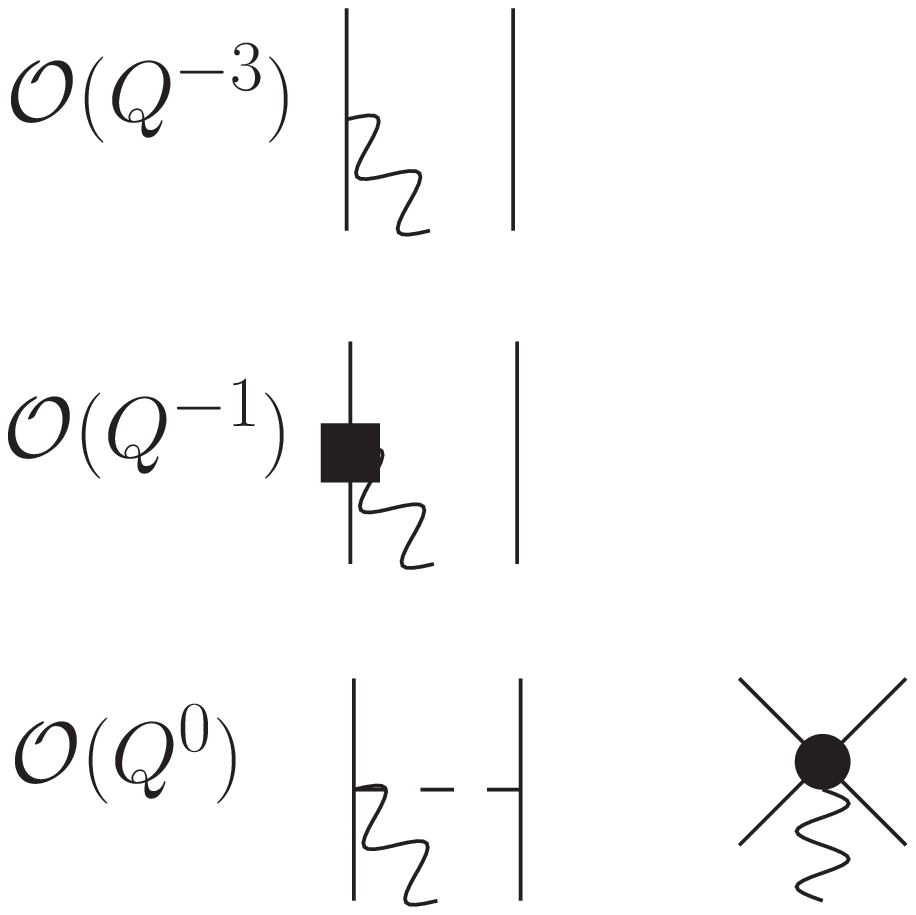}
\caption{On the left: diagrams illustrating one- and two-body 
$\chi$EFT EM currents 
entering at LO ($Q^{-2}$), NLO ($Q^{-1}$), 
N2LO ($Q^{0}$), and N3LO ($Q^{1}$). 
On the right: diagrams illustrating one- and two-body $\chi$EFT axial
currents 
entering at LO ($Q^{-3}$), NLO ($Q^{-1}$), 
and N2LO ($Q^{0}$).
Nucleons, pions, and EW probes 
are denoted by solid, dashed, and wavy lines,
respectively. The solid square represents the relativistic corrections
to the one-body current, while the solid 
circles represent the contact terms. Only the relevant topologies
are indicated. Loop corrections to short range EM currents turn out to 
vanish. No contribution of order $Q^{-2}$ exists for the axial current.}
\label{fig:em+ax}       
\end{figure}

\paragraph{Electromagnetic current operator.}
The EM current operators are diagrammatically 
represented in the left panel of Fig.~\ref{fig:em+ax}, where
they are listed according to their scaling in $Q$. 
The leading order (LO) contribution consists of the well known single-nucleon
convection and magnetization currents, and is of order $Q^{-2}$, 
while the next-to-next-to leading order
(N2LO) contribution arises from the $(Q/m)^2$ relativistic corrections to the
previous contribution ($m$ is the nucleon mass), and is therefore of order 
$Q^{0}$.
The next-to-leading order (NLO) term involves seagull and 
in-flight long-range contributions associated with one-pion exchange (OPE). 
The next-to-next-to-next-to-leading order (N3LO) currents, therefore at order
$Q^1$, 
consist of (i) one-loop two-pion-exchange (TPE) terms, 
(ii) OPE terms induced by $\gamma\pi N$ interactions beyond LO,
and (iii) contact terms generated by
minimal substitution in the four-nucleon contact interactions
involving two gradients of the nucleon fields, as well as by
nonminimal couplings to the electromagnetic field. 
The former are linked to the $\chi$EFT potential at order $Q^2$ 
via current conservation, and therefore they involve the same
LECs entering the $\chi$EFT NN interaction. These are taken from 
fits to the NN scattering data.
On the other hand, the LECs entering the nonminimal contact currents
as well as those entering the N3LO OPE contribution 
need to be fixed to EM observables. 
The explicit expression for all these N3LO currents can be found
in Ref.~\cite{Pia13}. Note that
the two-body $\chi$EFT operators have
a power-law behavior at large momenta, which requires
a regularization procedure. This is implemented via
the introduction of a cutoff function of the form exp(-$Q^4/\Lambda^4$), 
where $\Lambda$ = 500 or 600 MeV (the same values of 
the NN potential and TNI). 

We consider in some detail only those N3LO contributions 
involving new LECs,
i.e., the nonminimal and OPE currents. These terms can be written as
\begin{eqnarray}
{\bf J}_{\rm nm}(ij)&\propto & {\bf q}\times[{{d_1^S}}{\bm\sigma}_i + 
{{d_1^V}} (\tau_{i}^z-\tau_{j}^z){\bm\sigma}_i] + i\rightleftharpoons j
\label{eq:j_nm} \\
{\bf J}_{\rm OPE}(ij)&\propto & 
\frac{{\bm\sigma}_j\cdot{\bf k}_j}{(m_\pi^2+{\bf k}_j^2)}
\,{\bf q}\times[({{d_2^S}}{\bm\tau}_i\cdot{\bm\tau}_j\,+{{d_2^V}}\tau_{j}^z)
{\bf k}_j +{{d_3^V}}({\bm \tau}_i\times{\bm\tau}_j)^z
{\bm\sigma}_i\times{\bf k}_j] + i\rightleftharpoons j\ , \label{eq:j_OPE}
\end{eqnarray}
where ${\bf q}$ is the photon momentum, ${\bm\sigma}_i$ (${\bm\tau}_i$)
are the spin (isospin) Pauli matrices, ${\bf k}_i$ is the momentum transfer
to nucleon $i$. The LECs $d_1^S$, $d_1^V$, $d_2^S$, $d_2^V$ and $d_3^V$ need
to be fixed on EM observables.
The adopted fitting procedure is extensively discussed in Ref.~\cite{Pia13}.
Here we only summarize the main features: (i) the LECs multiplying
isoscalar operators ($d_1^S$ and $d_2^S$) are fixed so as to reproduce
the deuteron magnetic moment and the isoscalar combination of the $A=3$
magnetic moments ($\mu_S$). (ii) In order to achieve ``natural'' values for the 
LECs multiplying isovector operators and not to spoil chiral convergence,
two LECs 
($d_2^V$ and $d_3^V$) have been fixed by saturating the 
$\Delta$-resonance (a common strategy adopted in the 
literature),
and $d_1^V$ has been fixed by fitting either the
cross section for neutron-proton radiative capture at thermal energies,
$\sigma_{np}$,
or the isovector combination of the $A=3$ magnetic moments,
$\mu_V$. These two sets of LECs are called SET II and SET III, respectively. 
The values
for the different LECs are given in Ref.~\cite{Pia13}, where it
has been shown that the experimental value for 
$\sigma_{np}$ ($\mu_V$) is reproduced within 
few percent with the LECs of SET III (SET II).

Finally, we would like to remark that the most significant differences
between the model for the EM current 
presented here and that of Park {\it et al.}
arise at N3LO for the box diagrams and the contact terms. 
In particular, the contact terms of Park {\it et al.} are much simpler
than those presented above and can be written as sum of two
terms, one isoscalar and one isovector, with two different LECs in front.
These LECs have been fitted in Refs.~\cite{Mar12,Mar13} to reproduce 
$\mu_S$ and $\mu_V$.  
For a more detailed discussion of this point, see Ref.~\cite{Pastore}.

\paragraph{Weak axial current operator.}
The weak axial current operators are diagrammatically 
represented in the right panel of Fig.~\ref{fig:em+ax}, 
where, as in the EM case,
they are listed according to their scaling in $Q$. 
The LO contribution consists of the well known single-nucleon axial
current, and is of order $Q^{-3}$. At order $Q^{-2}$ it turns out that
there are no contributions, and therefore the NLO contribution is
of order $Q^1$, and arises from the $(Q/m)^2$ relativistic corrections to the
LO contribution.
The N2LO currents
consist of the OPE term and a contact term. Note that N3LO contributions
arise from loop and TPE terms, and they have not been calculated yet.
They are not illustrated in Fig.~\ref{fig:em+ax}.
The only model available for the axial current is that of 
Park {\it et al.}~\cite{Par03}, up to N2LO, with only one LEC, $d_R$.
As first shown in Ref.~\cite{Gar06}, such LEC can be related to the LEC
$c_D$ entering one of the two contact terms present in the TNI
at N2LO, via the relation
$ d_R=\frac{m}{\Lambda_\chi g_A} {c_D} + 
\frac{1}{3}m ({c_3}+2 {c_4})+\frac{1}{6}$,
%
where $g_A$ is the single-nucleon axial coupling constant, $c_3$ and
$c_4$ are LECs  of the $\pi N$ Lagrangian, already part of the 
chiral NN potential at NLO, and $\Lambda_\chi=700$ MeV is the 
the chiral-symmetry-breaking scale. 
Therefore, it has become common practice
to fit $c_D$ (and $c_E$--- the other LEC entering the N2LO TNI)
to the triton binding energy and the Gamow-Teller
matrix element in tritium $\beta$-decay. 
The values obtained in this way 
for $c_D$ and $c_E$ are listed in Ref.~\cite{Mar12} and they have been used
in Refs.~\cite{Mar12,Mar13} to study the muon capture on deuteron 
and $^3$He and the $pp$ capture, 
as it will be reviewed in Sec.~\ref{sec:res}. 
Note that the first studies of $A=3$ and 4
elastic scattering observables, as cross sections and analyzing powers, 
with these values for $c_E$ and $c_D$ have been reported in Ref.~\cite{Viv13}. 
A final remark: the model of Park {\it et al.} is the only one available 
at present for both the axial and the vector (or EM) current
operators. We have used this model in our studies of weak 
observables~\cite{Mar12,Mar13}. However, the 
significant differences found between the Park {\it et al.} model
and the most recent ones of Pastore {\it et al.} (and K\"olling {\it et al.})
in the EM sector make it urgent to perform a new derivation of the 
weak axial operators within TOPT, following the 
footsteps of Pastore {\it et al.} for the EM case. 

\section{Results}
\label{sec:res}

In this section we first present some selected results on the electromagnetic
structure of the deuteron, the triton and $^3$He. A complete discussion 
can be found in Ref.~\cite{Pia13}. Then we present results 
for the muon capture on 
deuteron and $^3$He~\cite{Mar12}, 
and for the $pp$ reaction~\cite{Mar13}, respectively.

\paragraph{The electromagnetic structure of $A=2,3$ nuclei.}
The static properties of $A=2,3$ nuclei are summarized in 
Table~\ref{tab:static}, where we present the $\chi$EFT results for the
deuteron r.m.s. radius and quadrupole moment, and the charge
and magnetic radii for the $A=3$ nuclei. The experimental data are
also reported. Note that
the deuteron and $A=3$ magnetic moments are
used to fit the LECs
(in SET III). The theoretical uncertainties are due to the cutoff dependence
and the fitting procedure.
By inspection of the table, we can conclude that the
static properties of the $A=2,3$ nuclei are nicely reproduced.
It should be noticed that within the phenomenological approach,
based on the Argonne $v_{18}$ potential~\cite{Wir95} (AV18), 
the quadrupole moment
is calculated to be 0.275 fm$^{2}$, in significant disagreement 
with the experimental value.
\begin{table}[t]
\caption{Deuteron r.m.s. radius ($r_d$) and quadrupole moment ($Q_d$), and
$^3$H and $^3$He charge ($r_c$) and magnetic ($r_m$) radii. 
The corresponding
experimental values are also reported.}
\centering
\label{tab:static}       
\begin{tabular}{lll}
\hline\noalign{\smallskip}
 & Theory & Experiment  \\[3pt]
\tableheadseprule\noalign{\smallskip}
$r_d$ [fm] & 1.972 $\pm$ 0.004 & 1.9733 $\pm$ 0.0044 \\
$Q_d$ [fm$^2$] & 0.2836 $\pm$ 0.0016  & 0.2859 $\pm$ 0.0003 \\
$r_c(^3{\rm He})$ [fm] & 1.962 $\pm 0.004$  & 1.959 $\pm 0.030$ \\
$r_c(^3{\rm H})$ [fm]  & 1.756 $\pm 0.006$  & 1.755 $\pm 0.086$ \\
$r_m(^3{\rm He})$ [fm] & 1.905 $\pm 0.022$  & 1.965 $\pm 0.153$ \\
$r_m(^3{\rm H})$ [fm]  & 1.791 $\pm 0.018$  & 1.840 $\pm 0.181$ \\
\noalign{\smallskip}\hline
\end{tabular}
\end{table}

\begin{figure}
\centering
\vspace*{0.3cm} 
  \includegraphics[width=0.25\textwidth]{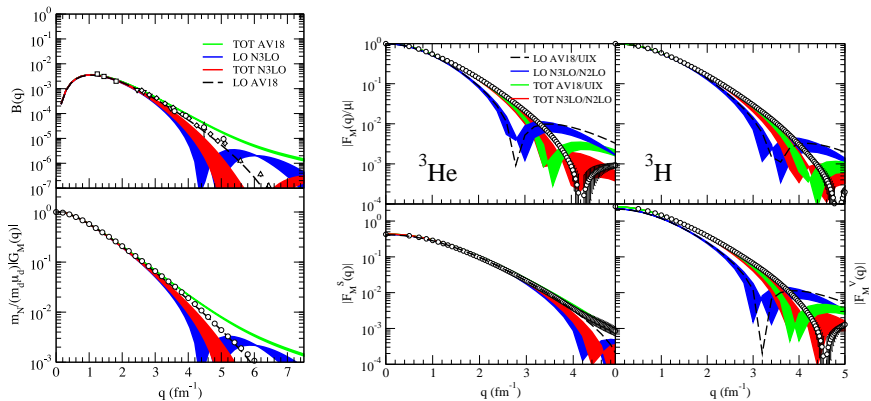}
\hspace*{0.2cm}
  \includegraphics[width=0.45\textwidth]{Current3body.eps}
\vspace*{0.6cm} 
\caption{The deuteron $B(q )$ structure function
and magnetic form factor (on the left), and the 
$A=3$ magnetic form factors, together with their isoscalar
and isovector combinations (on the right), 
obtained
at leading order (LO) and with inclusion of current operators
up to N3LO (TOT), compared with the experimental data.
The curves labelled ``AV18'' (or ``AV18/UIX'') 
have been obtained
within the hybrid $\chi$EFT approach using the AV18 (and 
Urbana IX~\protect\cite{Pud95} TNI for $A=3$)
phenomenological potentials.
The curves labelled ``N3LO'' (or ``N3LO/N2LO'') have been obtained
instead using the N3LO NN potential~\protect\cite{Ent03} 
(and N2LO~\protect\cite{Nav07} TNI for $A=3$).}
\label{fig:a2-3m}       
\end{figure}

The deuteron $B(q)$ structure function and magnetic form factor,
together with the $A=3$ magnetic form factors,
and their isoscalar and isovector combinations, are given 
in Fig.~\ref{fig:a2-3m}. By inspection of the figure, we can conclude that
the $\chi$EFT calculation is in agreement with the experimental data
in a range of $q$-values much larger than one would naively expect,
(up to 1-2 times the pion mass). On the other hand, this $\chi$EFT
calculation is unable to reproduce the first diffraction region of the
$A=3$ magnetic form factors, a problem which is present also in the 
hybrid calculation, as well as in the phenomenological one
(see for instance Ref.~\cite{Mar98-05}).

\paragraph{Muon capture on $A=2,3$ nuclei.}
The muon capture reactions here under consideration are
$\mu^-+d\rightarrow n+n+\nu_\mu$ and 
$\mu^-+\,^3{\rm He}\rightarrow \,^3{\rm H} +\nu_\mu$, 
for which we are interested in the capture rate in the doublet
hyperfine initial state ($\Gamma^D$) 
and in the total capture rate ($\Gamma_0$), respectively.
The $\chi$EFT results of Ref.~\cite{Mar12}
are summarized in Table~\ref{tab:muon},
where the value for $\Gamma^D$ obtained retaining in the final $nn$ scattering
state only the $^1S_0$ partial wave is also shown. The results of the
table can be summarized as $\Gamma^D=(399\pm 3)$ sec$^{-1}$ and
$\Gamma_0=(1494 \pm 21)$ sec$^{-1}$. The errors are due to 
cutoff dependence, the uncertainty inherent  
the fitting procedure of the LEC $d_R$, and radiative 
corrections~\cite{Mar12}. The experimental data for $\Gamma^D$
are affected at present by quite large uncertainties 
and no significant comparison between theory and experiment can be
made. Instead, the theoretical prediction for $\Gamma_0$ is 
in very nice agreement with the experimental determination of
Ref.~\cite{Ack98}, $(1496\pm 4)$ sec$^{-1}$. 

\begin{table}[t]
\caption{Doublet capture rate for muon capture on deuteron and total
capture rate for muon capture on $^3$He
obtained
at leading order (LO) and with inclusion of current operators
up to N2LO (TOT). The results calculated with the different values of two
cutoff $\Lambda$ are reported. The 
theoretical uncertainties are due to the fitting procedure of the LEC $d_R$.
}
\centering
\label{tab:muon}       
\begin{tabular}{llll}
\hline\noalign{\smallskip}
 & $\Gamma^D(^1S_0)$ [sec$^{-1}$] & $\Gamma^D$ [sec$^{-1}$] & $\Gamma_0$ [sec$^{-1}$]\\
[3pt]
\tableheadseprule\noalign{\smallskip}
LO - $\Lambda=500$ MeV & 238.8 & 381.7 & 1362 \\
LO - $\Lambda=600$ MeV & 238.7 & 380.8 & 1360 \\
TOT - $\Lambda=500$ MeV & 254.4$\pm 9$ & 399.2$\pm 9$ & 1488$\pm 9$ \\
TOT - $\Lambda=600$ MeV & 255$\pm 1$ & 399$\pm 1$ & 1499$\pm 9$ \\
\noalign{\smallskip}\hline
\end{tabular}
\end{table}

\begin{figure}
\centering
\vspace*{0.4cm}
  \includegraphics[width=0.38\textwidth]{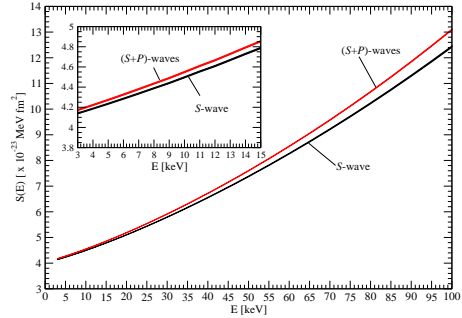}
\caption{The astrophysical $S$-factor for 
$E=3-100$ keV.  The $S$- and
$(S+P)$-wave contributions are displayed separately.  In
the inset, $S(E)$ is shown in the range 3--15 keV.}
\label{fig:se}       
\end{figure}

\paragraph{Weak proton-proton capture.}
The astrophysical $S$-factor for $pp$ weak
capture 
is typically given as a Taylor expansion around
$E=0$, and the coefficients
of the expansion, $S(0)$, $S'(0)$ and $S''(0)$, are the quantities of
interest~\cite{SFII}. Alternatively, the energy dependence of $S(E)$
can be made explicit by calculating it directly. Note that the
Gamow peak for the $pp$ reaction is at $E=6$ keV in the Sun, 
and it becomes of about 15 keV in larger mass stars. Therefore, we have
studied the $pp$ reaction in the energy range $E=3-100$ keV.
Two ingredients are essential in the calculation: (i) the initial
$pp$ scattering state is calculated using the $\chi$EFT N3LO 
potential~\cite{Ent03}
augmented not only of the Coulomb interaction, but also of the
higher order electromagnetic terms, due
to two-photon exchange and vacuum polarization.
The additional
distortion of the $pp$ wave function, induced primarily by
vacuum polarization, has been shown to reduce $S(0)$ by
$\sim 1$\% in previous studies (see Ref.~\cite{SFII} and references
therein).
(ii) To have the correct energy-dependence of the $S$-factor
up to $E=100$ keV, we have included, in addition to the $S$-wave
(the $^1S_0$ partial wave), all the $P$-wave channels
($^3P_0$, $^3P_1$, and $^3P_2$), and we have retained the explicit
dependence on the momentum transfer
${\bf q}={\bf p}_e+{\bf p}_\nu$ (${\bf p}_e$ and ${\bf p}_\nu$ are the electron
and neutrino momenta, respectively) via a standard multipole
expansion. More details can be found in Ref.~\cite{Mar13}.
Finally, we recall that the model for the weak current is the one 
discussed in Sec.~\ref{sec:ew_op}, and is the same used in the 
successful studies 
of the muon capture reactions presented above.
The $S$-factor at zero energy is found to be $S(0)=(4.030
\pm 0.006)\times 10^{-23}$ MeV fm$^2$,
with a $P$-wave contribution of $0.020\times 10^{-23}$
MeV fm$^2$.  The theoretical uncertainty is due to the fitting
procedure of the LECs and to the cutoff dependence. This value is
$\simeq 1$\% larger than the value reported in the literature~\cite{SFII}.
The energy dependence of $S(E)$ is shown in Fig.~\ref{fig:se}.
The $S$- and $(S+P)$-wave contributions are displayed separately, and
the theoretical uncertainty is included---the curves are in fact very
narrow bands.  As expected, the $P$-wave contributions become
significant at higher values of $E$. From these results, with a least
$\chi^2$ fitting procedure, we have calculated the coefficients 
$S'(0)$ and $S''(0)$, which are listed in Table II of 
Ref.~\cite{Mar13}, where a thorough discussion of these results
is also present.

\section{Conclusions and outlook}
\label{sec:concl}
We have presented the most recent studies of
EW observables involving light nuclei within the $\chi$EFT
framework. The considered observables are in good agreement with the
available experimental data, except for the $A=3$ magnetic form factors
at high values of momentum transfer, in a region, though, 
well beyond the applicability range of $\chi$EFT. To be noticed that,
while the EM observables have been studied with the most recent
model for the $\chi$EFT EM operators up to N3LO, 
the weak observables have been studied
with the model of Park {\it et al.} up to N2LO. 
Therefore, it is highly desirable to derive the axial currents
up to N3LO in TOPT 
and to repeat the studies for the muon captures and $pp$ reaction.
Finally, it should be noticed that
the radiative proton-deuteron and the weak proton-$^3$He captures,
also relevant in astrophysics,
are at reach within the present framework. Work along these lines
is currently underway.


\begin{acknowledgements}
The Author would like to thank the Gran Sasso Science Institute (INFN),
and especially Prof.\ F.\ Vissani, for the support and warm hospitality
extended to her in occasion of her visit in the Fall 2013, during which
this work was completed. 
\end{acknowledgements}



\end{document}